\documentclass[12pt]{iopart}

\usepackage{graphicx}
\usepackage{subfigure}
\usepackage{verbatim}
\usepackage{dcolumn}
\usepackage{bm}
\usepackage{epsf}
\usepackage{color}
\usepackage[colorlinks=true,citecolor=blue,linkcolor=blue,urlcolor=blue]{hyperref}
\usepackage{hhline}
\usepackage{amssymb}
\usepackage{cite}
\usepackage{tikz}
\usetikzlibrary{arrows,shapes,calc,matrix}

  \expandafter\let\csname equation*\endcsname\relax
  \expandafter\let\csname endequation*\endcsname\relax
\usepackage{amsmath}

\makeatletter
\newcommand\footnoteref[1]{\protected@xdef\@thefnmark{\ref{#1}}\@footnotemark}
\makeatother

\newcommand{\bra}[1]{\left\langle #1\right|}
\newcommand{\ket}[1]{\left|#1\right\rangle}

\newcommand{\pd}{\partial}

\newcommand{\ex}[1]{\exp{\left(#1\right)}}
\newcommand{\loge}[1]{\ln{\left(#1\right)}}
\newcommand{\id}{\mathbb{I}}

\newcommand{\bla}{bla\\bla\\bla\\bla\\bla}

\newcommand{\mc}[1]{\mathcal{#1}}

\newcommand{\mf}[1]{\mathfrak{#1}}
\newcommand{\mrm}[1]{\mathrm{#1}}

\newcommand{\draftmode}{1}    
\newcommand{\notetoself}[1]{\ifnum \draftmode=1 {\color[rgb]{0,0,0.8} [#1]} \fi}  
\newcommand{\cuttext}[1]{\ifnum \draftmode=1 {\color[rgb]{0,0.5,0} [#1]} \fi}  
\newcommand{\warntext}[1]{\ifnum \draftmode=1 {\color[rgb]{0.9,0.6,0} #1} \else {#1} \color{black} \fi}

\begin{document}

\title{Foundations of statistical mechanics from symmetries of entanglement}


\author{Sebastian Deffner$^{1,2}$ and Wojciech H. Zurek$^1$}
\address{$^1$Theoretical Division, Los Alamos National Laboratory, Los Alamos, NM 87545, USA}
\address{$^2$Center for Nonlinear Studies, Los Alamos National Laboratory, Los Alamos, NM 87545, USA}

\date{}

\begin{abstract}
Envariance -- entanglement assisted invariance -- is a recently discovered symmetry of composite quantum systems. We show that thermodynamic equilibrium states are fully characterized by their envariance. In particular, the microcanonical equilibrium  of a system $\mc{S} $ with Hamiltonian $H_\mc{S}$ is a fully energetically degenerate quantum state envariant under every unitary transformation. The representation of the canonical equilibrium then follows from simply counting degenerate energy states. Our conceptually novel approach is free of mathematically ambiguous notions such as ensemble, randomness,  etc., and, while it does not even rely on probability, it helps to understand its role in the quantum world. 
\end{abstract}

\section{Introduction}
Statistical physics was developed in the XIX century. The fundamental physical theory was then Newtonian mechanics. The key task of statistical physics was to bridge the chasm between  microstates (points in phase space) and thermodynamic macrostates (given by temperature, entropy, pressure, etc.) Strictly speaking, that chasm is -- within the context of Newtonian mechanics -- unbridgeable, as classical microstates have vanishing entropy. Thus, a ``half-way house" populated with fictitious but useful concepts such as \textit{ensembles} was erected -- half way between micro and macro -- and served as a pillar supporting the bridge. Even before ensembles were officially introduced by Gibbs \cite{Gibbs1902}, the concept was \emph{de facto} used by, e.g., Maxwell \cite{Maxwell1860} and Boltzmann \cite{Boltzmann1877}. Doubts about this ``half-way house" strategy nevertheless remained, as controversies surrounding the H-theorem demonstrate. 

Our point is that, while in XIX century physics ensembles were necessary because thermodynamics anticipated the role played by information in quantum physics, the state of a {\it single} quantum system can be mixed. Therefore, the contradiction between the pure classical microstate and an (impure) macrostate does not arise in a Universe that is quantum to the core. Yet, the development of quantum statistical physics consisted to a large extent of re-deploying strategies developed to deal with the fundamental contradiction between Newtonian physics and thermodynamics. 

We claim that such ``crutches'' that were devised (and helpful) in the XIX century became unnecessary with the advent of quantum physics in the XX century. Thus, in the XXI century we can simply dispose of the ensembles invented to justify the use of probabilities representing the ignorance of the observers and to compute the entropy of the macrostate.

In the present paper we propose an alternative approach to the foundations of statistical mechanics that is free of the conceptual caveats of classical theory, and relies purely on quantum mechanical notions such as entanglement. Our approach is based on \textit{envariance} (or entanglement assisted invariance) -- a symmetry-based view of probabilities that has been recently developed to \textit{derive} Born's rule from the non-controversial quantum postulates \cite{Zurek2003,Zurek2003a,Zurek2005,Zurek2009,Zurek2011} -- and that is, therefore, exceptionally well-suited to analyze probabilistic notions in quantum theories.

\subsection{Entanglement assisted invariance} Consider a quantum system, $\mc{S}$, which is maximally entangled with an environment, $\mc{E}$, and let $\ket{\psi_\mc{SE}}$ denote the composite state in $\mc{S}\otimes\mc{E}$. Then $\ket{\psi_\mc{SE}}$ is called envariant under a unitary map $U_\mc{S}=u_\mc{S}\otimes\id_\mc{E}$ iff there exists another unitary $U_\mc{E}=\id_\mc{S}\otimes u_\mc{E}$ such that,
\begin{equation}
\label{eq01}
U_\mc{S} \ket{\psi_\mc{SE}} =\left(u_\mc{S}\otimes\id_\mc{E}\right)\ket{\psi_\mc{SE}}=\ket{\eta_\mc{SE}};
\end{equation}
\vspace{-0.3in}
\begin{equation}
\label{eq02}
U_\mc{E}\ket{\eta_\mc{SE}}=\left(\id_\mc{S}\otimes u_\mc{E}\right)\ket{\eta_\mc{SE}}=\ket{\psi_\mc{SE}}\,.
\end{equation}
Thus, $U_\mc{E}$ that does not act on $\mc{S}$ ``does the job'' of the inverse map of $U_\mc{S}$ on $\mc{S}$ -- assisted by the environment $\mc{E}$.

The principle is most easily illustrated with a simple example. Suppose $\mc{S}$ and $\mc{E}$ are each given by two-level systems, where $\{\ket{\uparrow}_\mc{S}, \ket{\downarrow}_\mc{S}\}$ are the eigenstates of $\mc{S}$ and $\{\ket{\uparrow}_\mc{E}, \ket{\downarrow}_\mc{E}\}$ span $\mc{E}$. Now, further assume $\ket{\psi_\mc{SE}}\propto\ket{\uparrow}_\mc{S}\otimes\ket{\uparrow}_\mc{E}+\ket{\downarrow}_\mc{S}\otimes\ket{\downarrow}_\mc{E}$ and $U_\mc{S}$ is a \textit{swap} in $\mc{S}$ -- it ``flips'' its spin. Then, we have
\begin{equation}
\label{eq02a}
 \begin{tikzpicture}[>=stealth,baseline,anchor=base,inner sep=0pt]
      \matrix (foil) [matrix of math nodes,nodes={minimum height=0.5em}] {
         & {\color{blue}\ket{\uparrow}_\mc{S}} & \otimes & \ket{\uparrow}_\mc{E} &  & + &  & {\color{blue}\ket{\downarrow}_\mc{S}} & \otimes & \ket{\downarrow}_\mc{E} &  \xrightarrow{\quad {\color{blue}U_\mc{S}}\quad} 
{\color{blue}\ket{\downarrow}_\mc{S}}\otimes\ket{\uparrow}_\mc{E}+{\color{blue}\ket{\uparrow}_\mc{S}}\otimes\ket{\downarrow}_\mc{E}\,.\\
      };
      \path[->] ($(foil-1-2.north)+(0,1ex)$)   edge[blue,bend left=45]    ($(foil-1-8.north)+(0,1ex)$);
      \path[<-] ($(foil-1-2.south)-(0,1ex)$)   edge[blue,bend left=-45]    ($(foil-1-8.south)-(0,1ex)$);
    \end{tikzpicture}
\end{equation}
The action of $U_\mc{S}$ on $\ket{\psi}_\mc{SE}$ can be restored by a swap, $U_\mc{E}$, on $\mc{E}$,
  \begin{equation}
    \begin{tikzpicture}[>=stealth,baseline,anchor=base,inner sep=0pt]
      \matrix (foil) [matrix of math nodes,nodes={minimum height=0.5em}] {
         & \ket{\downarrow}_\mc{S} & \otimes & {\color{red}\ket{\uparrow}_\mc{E}} &  & + &  & \ket{\uparrow}_\mc{S} & \otimes & {\color{red}\ket{\downarrow}_\mc{E}} & \xrightarrow{\quad {\color{red}U_\mc{E}}\quad} 
\ket{\downarrow}_\mc{S}\otimes{\color{red}\ket{\downarrow}_\mc{E}}+\ket{\uparrow}_\mc{S}\otimes{\color{red}\ket{\uparrow}_\mc{E}}\,. \\
      };
      \path[->]  ($(foil-1-4.north)+(0,1ex)$) edge[red,bend left=45] ($(foil-1-10.north)+(0,1ex)$);
       \path[<-]  ($(foil-1-4.south)-(0,1ex)$) edge[red,bend left=-45] ($(foil-1-10.south)-(0,1ex)$);
    \end{tikzpicture}
  \end{equation}
Thus, the swap $U_\mc{E}$ on $\mc{E}$ restores the pre-swap $\ket{\psi}_\mc{SE}$ without ``touching'' $\mc{S}$, i.e., the global state is restored by solely acting on $\mc{E}$. Consequently, local probabilities of the two swapped spin states are both exchanged and unchanged. Hence, they have to be equal. This provides the fundamental connection of quantum states and probabilities \cite{Zurek2005}, and leads to Born's rule.

Envariance of pure states is a purely quantum symmetry. Classical pure states of composite systems are given by Cartesian rather then tensor products. Therefore, it can be shown that such an environment assisted inverse of a map acting on a classical system cannot exist \cite{Zurek2005}. This reflects the fact that some properties of quantum states can be considered ``relative'' with respect to $\mc{E}$, whereas all properties of classical states are always ``absolute''. 

Recent experiments \cite{Vermeyden2015,Karimi2016} have shown that envariance is not only a theoretical concept, but a physical reality. Thus, in the present work we will use envariance to advance a conceptually novel approach to the foundations of statistical mechanics. 

\section{Microcanonical state from envariance}

We begin by considering the microcanonical equilibrium. Generally, thermodynamic equilibrium states are characterized by extrema of physical properties, such as maximal phase space volume, maximal thermodynamic entropy, or maximal randomness \cite{Uffink2007}.  We will define the microcanonical equilibrium as the quantum state that is ``maximally envariant'', i.e., envariant under all unitary operations on $\mc{S}$. To this end, we write the composite state $\ket{\psi_\mc{SE}}$ in Schmidt decomposition,
\begin{equation}
\label{eq03}
\ket{\psi_\mc{SE}}=\sum_k a_k \ket{s_k}\otimes\ket{\varepsilon_k}\,,
\end{equation}
where by definition $\{\ket{s_k}\}$ and $\{\ket{\varepsilon_k}\}$ are orthocomplete in $\mc{S}$ and $\mc{E}$, respectively. The task is now to identify the ``special'' state that is maximally envariant.

It has been shown \cite{Zurek2005} that $\ket{\psi_\mc{SE}}$ is envariant under all unitary operations if and only if the Schmidt decomposition is even, i.e., all coefficients have the same absolute value, $\left|a_k\right|=\left|a_l\right|$ for all $l$ and $k$. We then can write,
\begin{equation}
\label{eq04}
\ket{\psi_\mc{SE}}\propto \sum_k \ex{i\phi_k} \ket{s_k}\otimes\ket{\varepsilon_k}\,,
\end{equation}
where $\phi_k$ are phases. Recall that in classical statistical mechanics equilibrium ensembles are identified as the states with the largest corresponding volume in phase space \cite{Uffink2007}. In the present context this ``identification'' readily translates into an equilibrium state that is envariant under the maximal number of, i.e.,  \emph{all}  unitary operations. 

It is interesting to note that Eq.~\eqref{eq04}  can also be used to derive Born's rule \cite{Zurek2003,Zurek2003a}. In particular, it can be shown that an outside observer would measure each of the states $\ket{s_k}$ with equal probability.  However, we emphasize once again that the present analysis is independent of any notion of probability, and all steps rely purely on symmetries of entanglement.

\newpage

To conclude the derivation we note that the microcanonical state is commonly identified as the state that is also fully energetically degenerate\footnote{\label{note1}We would like to emphasize that this is a standard assumption in textbook derivations \cite{Callen1985}. It can be easily relaxed (also employing standard tools) by considering an energy interval \cite{Toda1983}. For related discussions in a context similar to the present work see also Refs.~\cite{Popescu2006,Goldstein2006}.}  \cite{Callen1985}. To this end,  denote the Hamiltonian of the composite system by
\begin{equation}
\label{eq07}
H_\mc{SE}=H_\mc{S}\otimes \id_\mc{E}+\id_\mc{S}\otimes H_\mc{E}\,.
\end{equation}
Then, the internal energy of $\mc{S}$ is given by the quantum mechanical average
\begin{equation}
\label{eq08}
E_\mc{S}=\bra{\psi_\mc{SE}} \left(H_\mc{S}\otimes \id_\mc{E}\right)\ket{\psi_\mc{SE}}=\sum_k \bra{s_k} H_\mc{S}\ket{s_k}/Z\,,
\end{equation} 
where $Z$ is the energy-dependent dimension of the Hilbert space of $\mc{S}$, which is commonly also called the microcanonical partition function \cite{Callen1985}. Since $\ket{\psi_\mc{SE}}$ \eqref{eq04} is envariant under all unitary maps we can assume without loss of generality\footnoteref{note1} that $\{s_k\}_{k=1}^Z$ is a representation of the energy eigenbasis corresponding to $H_\mc{S}$, and we have $  \bra{s_k} H_\mc{S}\ket{s_k}=e_k$ with $E_\mc{S}=e_k=e_{k'}$ for all $k,k'\in\{1,\dots,Z\}$.

In conclusion, we have identified the fully quantum mechanical representation of the microcanonical state by two conditions\footnote{\label{note2}\textit{Note} that in our framework the microcanonical equilibrium is not represented by a unique state, but rather by an equivalence class of all maximally envariant states with the same energy.}: the state representing the microcanonical equilibrium of a system $\mc{S} $ with Hamiltonian $H_\mc{S}$ is the state that is (i) envariant under all unitary operations on $\mc{S}$ and (ii) fully energetically degenerate with respect to $H_\mc{S}$.

\section{Reformulation of the fundamental statement}

Before we continue to rebuild the foundations of statistical mechanics using envariance, let us briefly summarize and highlight what we have achieved so far. All previous treatments of the microcanonical state relied on notions such as probability, ergodicity, ensemble, randomness, indifference, etc. However, in the context of statistical physics none of these expressions are fully well-defined. Indeed,  in the early days of statistical physics seminal researchers such as Maxwell and Boltzmann struggled with the conceptual difficulties \cite{Uffink2007}. Modern interpretation and understanding of statistical mechanics, however, was invented by Gibbs, who simply ignored such foundational issues and made full use of the concept of probability.

In contrast, in our approach we only need a quantum symmetry induced by entanglement -- envariance -- instead of relying on mathematically ambiguous concepts. Thus, we can reformulate the fundamental statement of statistical  mechanics in quantum physics:
\begin{quote}
\textit{The microcanonical equilibrium of a system $\mc{S}$ with Hamiltonian $H_\mc{S}$ is a fully energetically degenerate quantum state envariant under all unitaries.}
\end{quote}
The remainder of this analysis will further illustrate this novel conceptual approach to the foundations of statistical mechanics by also treating the canonical equilibrium. 

\section{Canonical state from quantum envariance}

Let us now imagine that we can separate the total system $\mc{S}$ into a smaller subsystem of interest $\mf{S}$ and its complement, which we call heat bath $\mf{B}$. The Hamiltonian of $\mc{S}$ can then be written as 
\begin{equation}
\label{eq10}
H_\mc{S}=H_\mf{S}\otimes\id_\mf{B}+\id_\mf{S}\otimes H_\mf{B}+h_\mf{S, B}\,,
\end{equation}
where $h_\mf{S, B}$ denotes an interaction term. Physically this term is necessary to facilitate exchange of energy between the $\mf{S}$ and the heat bath $\mf{B}$. In the following, however, we will assume that $h_\mf{S, B}$ is sufficiently small so that we can neglect its contribution to the total energy, $E_\mc{S}=E_\mf{S}+E_\mf{B}$, and its effect on the composite equilibrium state  $\ket{\psi_\mc{SE}}$. These assumptions are in complete analogy to the ones of classical statistical mechanics \cite{Toda1983,Callen1985}. They will, however, be relaxed in a final part of the analysis.

Under these assumptions every composite energy eigenstate $\ket{s_k}$ can be written as a product,
\begin{equation}
\label{eq11}
\ket{s_k}= \ket{\mf{s}_k}\otimes\ket{\mf{b}_k}\,,
\end{equation}
where the states $ \ket{\mf{s}_k}$ and $\ket{\mf{b}_k}$ are energy eigenstates in $\mf{S}$ and $\mf{B}$, respectively. At this point envariance is crucial in our treatment: All orthonormal bases are equivalent under envariance\footnoteref{note2}. Therefore, we can choose $\ket{s_k}$ as energy eigenstates of $H_\mc{S}$.

For the canonical formalism we are now interested in the number of states accessible to the total system $\mc{S}$ under the condition that the total internal energy $E_\mc{S}$ \eqref{eq08} is given and constant.  When the subsystem of interest, $\mf{S}$, happens to be in a particular energy eigenstate $\ket{\mf{s}_k}$ then the internal energy of subsystem is given by the corresponding energy eigenvalue $\mf{e}_k$. Therefore, for the total energy $E_\mc{S}$ to be constant, the energy of the heat bath, $E_\mf{B}$, has to obey,
\begin{equation}
\label{eq13}
E_\mf{B}\left(\mf{e}_k\right)=E_\mc{S}-\mf{e}_k\,.
\end{equation}
This condition can only be met if the energy spectrum of the heat reservoir is at least as dense as the one of the subsystem. This observation will be illustrated shortly in Eqs.~\eqref{q16} and \eqref{q18}.

The number of states, $\mf{N}\left(\mf{e}_k\right)$, accessible to $\mc{S}$ is then given by the fraction
\begin{equation}
\label{eq14}
\mf{N}\left(\mf{e}_k\right)=\frac{\mf{N}_\mf{B}\left(E_\mc{S}-\mf{e}_k\right)}{\mf{N}_\mc{S}(E_\mc{S})}\,,
\end{equation} 
where $\mf{N}_\mc{S}(E_\mc{S}) $ is the total number of states in $\mc{S}$  consistent with Eq.~\eqref{eq08}, and $\mf{N}_\mf{B}\left(E_\mc{S}-\mf{e}_k\right)$ is the number of states available to the heat bath, $\mf{B}$, determined by condition \eqref{eq13}. In other words, we are asking for nothing else but the degeneracy in $\mf{B}$ corresponding to a particular energy state of the system of interest $\ket{\mf{s}_k}$.

\subsection{\label{sec:example}Example: Composition of multiple qubits}

The idea is most easily illustrated with a simple example, before we will derive the general formula in the following paragraph. Imagine a system of interest, $\mf{S}$, that interacts with $N$ non-interacting qubits with energy eigenstates $\ket{0}$ and $\ket{1}$ and corresponding eigenenergies $e^\mf{B}_0$ and $e^\mf{B}_1$.  Note once again that the composite states $\ket{s_k}$ can  be always chosen to be energy eigenstates, since the even composite state $\ket{\psi_\mc{SE}}$ \eqref{eq04} is envariant under all unitary operations on $\mc{S}$\footnoteref{note2}.  

We further assume the qubits to be non-interacting. Therefore, all energy eigenstates can be written in the form
\begin{equation}
\label{q15}
\ket{s_k}=\ket{\mf{s}_k}\otimes\underbrace{\ket{\delta^1_k \delta^2_k \cdots \delta^{N}_k}}_{N-qubits}\,.
\end{equation}
Here $\delta^i_k \in \{0,1\}$ for all $i\in {1,\dots,N}$  describing the states of the bath qubits. Let us   denote the number of qubits of $\mf{B}$ in $\ket{0}$ by $n$. Then the total internal energy $E_\mc{S}$ becomes a simple function of $n$ and is given by,
\begin{equation}
\label{q16}
E_\mc{S}=\mf{e}_k+n\, e^\mf{B}_0 + (N-n)\, e^\mf{B}_1\,.
\end{equation}
Now it is easy to see that the total number of states corresponding to a particular value of $E_\mc{S}$, i.e., the degeneracy in $\mf{B}$ corresponding to $\mf{e}_k$, \eqref{eq14} is given by,
\begin{equation}
\label{q17}
\mf{N}\left(\mf{e}_k\right)=\frac{N!}{n!\,(N-n)!}\,.
\end{equation} 
Equation~\eqref{q17} describes nothing else but the number of possibilities to distribute $n\, e^\mf{B}_0$ and $ (N-n)\, e^\mf{B}_1$ over $N$ qubits.

It is worth emphasizing that in the arguments leading to Eq.~\eqref{q17} we explicitly used that the $\ket{s_k}$ are energy eigenstates in $\mc{S}$ and the subsystem $\mf{S}$ and heat reservoir $\mf{B}$ are non-interacting \eqref{eq11}. The first condition is not an assumption, since the composite $\ket{\psi_{\mc{SE}}}$ is envariant under all unitary maps on $\mc{S}$, and the second condition  is in full agreement with conventional assumptions of thermodynamics.

\subsection{Boltzmann's formula for the canonical state}

The example treated in the preceding section can be easily generalized. We again assume that the heat reservoir $\mf{B}$ consists of $N$ non-interacting subsystems with identical eigenvalue spectra $\{e_j^\mf{B}\}_{j=1}^{m}$. This model includes the most relevant cases commonly analyzed in quantum thermodynamics, such as quantum Brownian motion \cite{Gemmer2009a}, where $\mf{B}$ consists of $N$ non-interacting harmonic oscillators.

In this case the internal energy \eqref{eq08} takes the form
\begin{equation}
\label{q18}
E_\mc{S}=\mf{e}_k+ n_1\, e_1^\mf{B} +n_2\, e_2^\mf{B}+\cdots +n_{m} e_{m}^\mf{B}\,,
\end{equation}
with $\sum_{j=1}^{m} n_j =N$. Therefore, the degeneracy \eqref{eq14} becomes
\begin{equation}
\label{q19}
\mf{N}\left(\mf{e}_k \right)=\frac{N!}{n_1! n_2!\cdots n_{m}!}\,.
\end{equation}
This expression is readily recognized as a quantum envariant formulation of Boltzmann's counting formula for the number of classical microstates \cite{Uffink2007}, which quantifies the volume of phase space occupied by the thermodynamic state. However, instead of having to equip  phase space with an (artificial) equispaced grid, we simply count degenerate states.

We are now ready to derive the Boltzmann-Gibbs formula. To this end consider that in the limit of very large, $N\gg 1$, $\mf{N}\left(\mf{e}_k \right)$ \eqref{q19} can be approximated with Stirling's formula. We have
\begin{equation}
\label{q20}
\loge{\mf{N}\left(\mf{e}_k \right)}= N\loge{N}-\sum_{j=1}^m n_j \loge{n_j}\,.
\end{equation}
As  pointed out earlier, thermodynamic equilibrium states are characterized by a maximum of symmetry or maximal number of ``involved energy states'', which corresponds classically to a maximal volume in phase space. In the case of the microcanonical equilibrium this condition was met by the state that is maximally envariant, namely envariant under all unitary maps. Now, following Boltzmann's line of thought we identify the canonical equilibrium by the configuration of the heat reservoir $\mf{B}$ for which the maximal number of energy eigenvalues are occupied. Under the constraints,
\begin{equation}
\label{q21}
\sum_{j=1}^{m} n_j =N\quad\text{and}\quad E_\mc{S}-\mf{e}_k=\sum_{j=1}^{m} n_j\,e_j^\mf{B}
\end{equation}
this problem can be solved by variational  calculus. One obtains
\begin{equation}
\label{q22}
n_j=\mu\,\ex{\lambda\,e_j^\mf{B}}\,,
\end{equation}
which is the celebrated Boltzmann-Gibbs formula. Notice that Eq.~\eqref{q22} is the number of states in the heat reservoir $\mf{B}$ with energy $e_j^\mf{B} $ for $\mf{S}$ and $\mf{B}$ being in thermodynamic, canonical equilibrium. In this treatment temperature merely enters through the Lagrangian multiplier $\lambda$.

What remains to be shown is that $\lambda$, indeed, characterizes the unique temperature of the system of interest, $\mf{S}$. To this end, imagine that the total system $\mc{S}$ can be separated into two small systems $\mf{S}_1$ and $\mf{S}_2$ of comparable size, and the thermal reservoir, $\mf{B}$. It is then easy to see that the total number of accessible states $\mf{N}\left(\mf{e}_k \right)$ does not significantly change in comparison to the previous case. In particular, in the limit of an infinitely large heat bath $\mf{B}$ the total number of accessible states for $\mf{B}$ is still given by Eq.~\eqref{q20}. In addition, it can be shown that the resulting value of the Lagarange multiplier, $\lambda$, is unique \cite{Wachsmuth2013a}. Hence, we can formulate a statement of the zeroth law of thermodynamics from envariance -- namely, two systems $\mf{S}_1$ and $\mf{S}_2$, that are in equilibrium with a large heat bath $\mf{B}$, are also in equilibrium with each other, and they have the same temperature corresponding to the unique value of $\lambda$.

The present discussion is exact, up to the approximation with the Stirling's formula, and only relies  on the fact that the total system $\mc{S}$ is in a microcanonical equilibrium as defined in terms of envariance \eqref{eq04}. The final derivation of the Boltzmann-Gibbs formula \eqref{q22}, however, requires additional thermodynamic conditions. In the case of the microcanonical equilibrium we replaced conventional arguments by maximal envariance, whereas for the canonical state  we required the maximal number of energy levels of the heat reservoir to be ``occupied''.

\section{Concluding remarks}

In thermodynamics equilibrium states are characterized by sets of canonical variables \cite{Callen1985}. For instance, for a closed system  of given volume and given number of particles the internal energy, $E$, the volume, $V$, and the number of particles, $N$, are sufficient to fully describe all thermodynamic properties.  These canonical variables are ``macroscopic'' in the sense that they are commonly interpreted as averages over many ``permissible microstates''. From a ``macroscopic'' point of view it appears plausible that there may exist a vast number of microstates consistent with specific values of $E$, $V$, and $N$. In classical statistical mechanics \cite{Callen1985} one then assumes ergodicity, i.e., that the individual system undergoes rapid, random transitions among these microstates, which is an excuse to ignore microstates and deal solely with the macrostate. This assumption is commonly formulated as the central \textit{postulate} of statistical mechanics stating that \cite{Callen1985}
\begin{quote}
\textit{a macroscopic system samples every permissible quantum state with equal probability.}
\end{quote}
Although this postulate appears to be plausible, and has proven to be powerful, it also poses several unanswered questions. For instance, it neither specifies the dynamics of the transitions nor does it specify the notion of probability that should be applied.

Nevertheless, such formulations of the fundamental postulate have been used also in quantum analyses, e.g., to prove ``Canonical Typicality'' \cite{Goldstein2006}, or to study the interplay of ``entanglement and the foundations of statistical mechanics'' \cite{Popescu2006,Deffner2015}. However, all these previous treatments had to rely to a certain extent on classical strategies that involved ensembles and in that context employed notions of probability and randomness. For instance, the starting point of the analysis in Ref.~\cite{Goldstein2006} is a set pure states, which is equipped with the Haar measure, i.e.,  an \textit{ad hoc} selected  probability distribution. Generally, however, probabilistic theories in physics are notoriously hard to formulate with mathematical rigor \cite{Uffink2007}. Even two centuries after Laplace's first description of the term ``probability'' \cite{Laplace1820} a conceptually universal approach to physical randomness -- e.g., the relation between Kolmogorov-like measure theory and the underlying physical states -- appears to be lacking within the classical context. Instead, several different notions of probability, such as subjective, objective, dispositional, frequentist etc. \cite{Muller2007,Uffink2009}, have been developed, all based on non-equivalent mathematical concepts.


The major achievement of the present analysis is a conceptually novel approach to the foundations to statistical mechanics. By exclusively relying on envariance -- entanglement assisted invariance -- we have identified quantum states that represent microcanonical equilibrium. The canonical equilibrium can   then be described by simple counting arguments of degenerate energy states in a heat bath. Such counting is reminiscent of classical statistical physics, but now we are counting not members of an ensemble, but provably equiprobable states. Our treatment is free of any mathematical or conceptual caveats. At no point ambiguous notions such as probability, ensemble, randomness, indifference, etc. have been employed.

\ack{This research was supported by the U.S. Department of Energy through a LANL Director's Funded Fellowship and LDRD, and, in part, by a grant from the Foundational Questions Institute Fund (fqxi.org) on the Physics of What Happens, grant \#2015-144057.}

\appendix

\section{Alternative treatment utilizing entropy} 
In the preceding analysis we re-formulated the foundations of statistical mechanics in terms of envariance. This appendix is dedicated to an alternative treatment of \eqref{eq14}, which involves the definition of the Boltzmann entropy $\mc{H}$. It is defined as the logarithm of the number of states,
\begin{equation}
\label{eq15}
\mc{H}\equiv k_\mrm{B} \loge{\mf{N}}
\end{equation}
where the parameter $k_\mrm{B}$ denotes the Boltzmann constant according to convention \cite{Callen1985}. With this definition Eq.~\eqref{eq14} can be re-written to read,
\begin{equation}
\label{eq16}
\begin{split}
&\mf{N}\left(\mf{e}_k\right)=\ex{\loge{\frac{\mf{N}_\mf{B}(E_\mc{S}-\mf{e}_k)}{\mf{N}_\mc{S}(E_\mc{S})}}}\\
&=\ex{k_\mrm{B}^{-1} \mc{H}_\mf{B}(E_\mc{S}-\mf{e}_k)-k_\mrm{B}^{-1} \mc{H}_\mc{S}(E_\mc{S})}\,.
\end{split}
\end{equation}
We continue our discussion by inspecting each term in the exponent of Eq.~\eqref{eq16} separately. The entropy of the total system $\mc{S}$ can be separated into a sum of the entropies of the subsystem of interest, $\mf{S}$, and the heat reservoir, $\mf{B}$, since we assume $\mf{S}$ and $\mf{B}$ to be uncorrelated. Accordingly, we have
\begin{equation}
\label{eq17}
\begin{split}
\mc{H}_\mc{S}(E_\mc{S})&=k_\mrm{B} \loge{\mf{N}_\mc{S}(E_\mc{S})}\\
&=k_\mrm{B} \loge{\mf{N}_\mf{S}(E_\mf{S}) \cdot \mf{N}_\mf{B}(E_\mf{B})}\\
&=\mc{H}_\mf{S}(E_\mf{S})+\mc{H}_\mf{B}(E_\mf{B})\,,
\end{split}
\end{equation}
where $E_\mf{S}$ and $E_\mf{B}$ denote the internal energies of subsystem and heat reservoir, respectively.

Now, we further assume that the internal energy of the heat reservoir is much larger than the one of the subsystem of interest, $E_\mf{B}\gg E_\mf{S}-\mf{e}_k$, which is justified for $\mrm{dim}(\mf{S})\ll \mrm{dim}(\mf{B})$ (see also Eq.~\eqref{q18}). Then, the first term in the exponent of Eq.~\eqref{eq16} can be written as
\begin{equation}
\label{eq18}
\begin{split}
\mc{H}_\mf{B}\left(E_\mc{S}-\mf{e}_k\right)\simeq\mc{H}_\mf{B}(E_\mf{B}) +\frac{\pd \mc{H}_\mf{B}}{\pd E_\mf{B}} \left(E_\mf{S}-\mf{e}_k\right)\,,
\end{split}
\end{equation}
where we identified $E_\mc{S}-E_\mf{S} \equiv E_\mf{B}$ and expanded the entropy of the heat reservoir around its average value. 

Substituting Eqs.~\eqref{eq17} and \eqref{eq18} into Eq.~\eqref{eq16} and making the usual thermodynamic identifications, $\pd \mc{H}_\mf{B}/\pd E_\mf{B} \equiv 1/T $ and $F_\mf{S}=E_\mf{S}-T\, \mc{H}_\mf{S}(E_\mf{S}) $, we obtain the canonical state,
\begin{equation}
\label{eq19}
\mf{N}\left(\mf{e}_k\right)=\ex{\frac{F_\mf{S}-\mf{e}_k}{k_\mrm{B} T} }\,.
\end{equation}
As before, the crucial steps in the derivation are identifying Eq.~\eqref{eq04} as microcanonical state and considering the fraction of states in Eq.~\eqref{eq14}. In this treatment, however, temperature is found to be a characteristic property of the heat bath.

We emphasize again that both treatments are pure expressions of envariance and are free of ambiguous notions such as probability, randomness, indifference, etc. The main difference between the  two presented derivations of the canonical equilibrium are: in the first treatment we obtained the canonical state for the heat reservoir $\mf{B}$, but we had to employ the additional thermodynamic assumption  of maximally many energy levels being occupied; the second approach yielded the canonical state for $\mf{S}$ with no further information about $\mf{B}$, and we had to introduce the concept of entropy \eqref{eq15}.

\subsection*{Canonical equilibrium with system-bath correlations}

Finally, we briefly analyze the situation of non-vanishing coupling Hamiltonians, $h_\mf{S,B}$. For the sake of simplicity we still assume that we can neglect its contribution to the internal energies, but we relax the assumption of $\ket{s_k}$ being a product state \eqref{eq11}. Generally, we have 
\begin{equation}
\label{eq20}
\ket{s_k}= \sum_\nu\gamma_\nu(k)\ket{\mf{s}_\nu(k)}\otimes\ket{\mf{b}_\nu(k)}\,,
\end{equation}
where $\ket{\mf{s}_\nu(k)}$ and $\ket{\mf{b}_\nu(k)}$ are again energy eigenstates in $\mf{S}$ and $\mf{B}$, respectively. This slight generalization allows for correlated and entangled $\mf{S}$ and $\mf{B}$.

As before we are interested in the fraction of states available to $\mc{S}$ under the condition that the total internal energy, $E_\mc{S}$, is given and constant \eqref{eq14}. In contrast to the previous case \eqref{eq17}, however, the total entropy $\mc{H}_\mc{S}(E_\mc{S})$ can no longer be written as a simple sum, since subsystem and heat reservoir are correlated. Instead we have 
\begin{equation}
\label{eq22}
\mc{H}_\mc{S}(E_\mc{S})=\mc{H}_\mf{S}(E_\mf{S})+\mc{H}_\mf{B}(E_\mf{B})+k_\mrm{B} \loge{\frac{\mf{N}_\mc{S}(E_\mf{S}+E_\mf{B})}{\mf{N}_\mf{S}(E_\mf{S})\cdot \mf{N}_\mf{B}(E_\mf{B})}}\,.
\end{equation}
Note that the Shannon information of a maximally mixed states \eqref{eq04} is identical to the logarithm of the dimension of the Hilbert space, i.e., here the number of microstates \cite{Ohya1983}. Thus, we can write with the reduced density matrices for the total system $\rho_\mc{S}$, the subsystem of interest $\rho_\mf{S}$, and the heat reservoir $\rho_\mf{B}$, 
\begin{equation}
\label{eq23}
D(\rho_\mc{S}||\rho_\mf{S}\otimes \rho_\mf{B})=\loge{\frac{\mf{N}_\mc{S}(E_\mf{S}+E_\mf{B})}{\mf{N}_\mf{S}(E_\mf{S})\cdot \mf{N}_\mf{B}(E_\mf{B})}}\,,
\end{equation}
where $D(\rho_\mc{S}||\rho_\mf{S}\otimes \rho_\mf{B})$ denotes the quantum mutual information \cite{Umegaki1954}.

Accordingly, the generalized canonical state accounting for correlation between subsystem and heat reservoir becomes,
\begin{equation}
\label{eq24}
\begin{split}
\mf{N}\left(\ket{\mf{s}_k}\right)&=\ex{\frac{F_\mf{S}-\mf{e}_k}{k_\mrm{B} T}-D(\rho_\mc{S}||\rho_\mf{S}\otimes \rho_\mf{B}) }\,.
\end{split}
\end{equation}
Equation~\eqref{eq24} generalizes the standard canonical state \eqref{eq19} in a natural and intuitive way as correlations between subsystem of interest and heat bath are quantified by the quantum mutual information \eqref{eq23}. 

\bibliographystyle{unsrt}
\bibliography{references_env}

\end{document}